\documentclass[twocolumn,showkeys,showpacs]{revtex4}
\usepackage{graphicx}
\usepackage{amssymb}
\usepackage{amsmath}

\newcommand{\ds}{\displaystyle}
\newcommand{\up}{\uparrow}
\newcommand{\dn}{\downarrow}
\newcommand{\de}{\partial}
\newcommand{\prll}{\parallel}

\begin{document}

\title{Intersubband spin-orbit coupling and spin splitting in symmetric quantum wells}

\author{F. V. Kyrychenko}
\author{C. A. Ullrich}
\affiliation{Department of Physics and Astronomy, University of Missouri, Columbia, Missouri, 65211, USA}
\author{I. D'Amico}
\affiliation{Department of Physics, University of York, York YO10 5DD, United Kingdom}

\begin{abstract}
In semiconductors with inversion asymmetry, spin-orbit coupling gives rise to
the well-known Dresselhaus and Rashba effects. If one considers quantum wells with two or more
conduction subbands, an additional, intersubband-induced spin-orbit term appears whose strength
is comparable to the Rashba coupling, and which remains finite for symmetric structures. We show that the
conduction band spin splitting due to this intersubband spin-orbit coupling term is negligible for
typical III-V quantum wells.
\end{abstract}


\maketitle

\section{Introduction}
Research in nanoscience is crucial for its technological implications and for the fundamental exploration
of the quantum properties of nanostructures such as quantum wells, wires and dots.
Of particular interest is the study of spin dynamics, which hopes to revolutionize traditional
electronics using the spin properties of the carriers (spintronics) \cite{book}.
In this context, the theoretical prediction \cite{Irene2000} and experimental confirmation \cite{Weber}
of the spin-Coulomb drag (SCD) effect  was of great importance, as this effect results
in the natural decay of spin current and  intrinsic dissipation in AC-spintronic circuits \cite{IC2006}.
Due to Coulomb interactions between spin-up and spin-down electrons,
the up and down components of the total linear momentum are not {\it separately} conserved.
This momentum exchange between the two populations represents an intrinsic
source of friction for spin currents, known as spin-transresistivity \cite{SCD}.

In \cite{IC2006} we demonstrated that the SCD produces an intrinsic linewidth in
spin-dependent optical excitations, which can be as big as a fraction of a meV for
intersubband spin plasmons in parabolic semiconductor quantum wells (QWs).
This intrinsic linewidth would be ideal to experimentally verify the behavior of the
spin-transresistivity in the frequency domain.

In our proposed experiment, we suggested to use {\it symmetric} parabolic QWs to avoid an
undesired splitting of the spin plasmons due to Rashba spin-orbit (SO) coupling.
We based our discussion on earlier work \cite{UllrichFlatte}, in which
collective intersubband spin excitations in QWs were described in the presence of
Dresselhaus and Rashba SO interaction terms \cite{dresselhaus,rashba}
for strictly two-dimensional (2D) systems \cite{winkler}.
In symmetric QWs, the Rashba term vanishes and only bulk inversion asymmetry (Dresselhaus)
interaction is present.

However, as shown recently by Bernardes et al. \cite{egues}, the Rashba SO coupling gives finite contributions
even for symmetric structures, if treated in higher order perturbation theory. As a consequence, for QWs with more than
one subband, there appears an additional intersubband SO interaction, whose magnitude can become comparable to that of
2D Dresselhaus and Rashba interactions. This interaction gives rise to a nonzero spin-Hall conductivity and
renormalizes the bulk mass by $\sim 5$\% in InSb double QWs \cite{egues}. This raises the question whether this effect
must be accounted for when extracting the SCD from intersubband spin plasmon linewidths \cite{IC2006}.

In this paper we are going to show that while intersubband SO interaction may manifest itself in some special cases,
as for example in the double well analyzed in Ref. \cite{egues},  it
has little to no effect on spin splitting and spin mixing in QWs once the 2D Dresselhaus and/or Rashba terms are
taken into account.

In Sec.~\ref{sec:gen} we present the general formalism of calculating conduction band states in quantum
structures including both 2D and intersubband SO interaction. In Sec.~\ref{sec:sym} we consider
the specific case of symmetric single-well quantum structures, and in Sec. \ref{parabolic} we present
results for a parabolic model QW. Sec. \ref{conclusions} gives a brief summary.

\section{General formalism}\label{sec:gen}

We consider conduction electrons in a QW described by the Hamiltonian
\begin{equation}\label{HH}
  \hat{H}=\hat{H}_0+\hat{H}_{\rm so},
\end{equation}
where $\hat{H}_0$ is spin independent and $\hat{H}_{\rm so}$ is the SO interaction
projected on the conduction band. For simplicity we will consider only spin off-diagonal (spin-mixing) terms in $\hat{H}_{\rm
so}$. The eigenfunctions associated with $\hat{H}_0$ alone can be obtained by solving a single-particle equation of the
Schr\"odinger-Poisson or Kohn-Sham type, resulting in spin-independent subband envelope functions $\psi_i(z,{\bf k}_{\prll})$
and energy eigenvalues $\varepsilon_i$,  where $i$ is the subband index and $z$ is the direction of quantum confinement.

Let us now consider the two lowest conduction subbands of the QW. In the basis of the first two subband spinors
$|\psi_1\up\rangle$,
$|\psi_1\dn\rangle$,
$|\psi_2\up\rangle$,
$|\psi_2\dn\rangle$,
the Schr\"{o}dinger equation with the full Hamiltonian (\ref{HH}) has the form
\begin{equation}\label{H}
  \left( \begin{array}{cccc}
    \varepsilon_1 & \alpha_1 & 0 & \beta \\
    \alpha_1^* & \varepsilon_1 & \beta' & 0 \\
    0 & \beta'^* & \varepsilon_2 & \alpha_2 \\
    \beta^* & 0 & \alpha_2^* & \varepsilon_2 \
  \end{array}\right) {\bf A}
=\varepsilon  {\bf A} ,
\end{equation}
where
\begin{eqnarray}
  \alpha_1&=&\langle\psi_1 \up |\hat{H}_{\rm so} |\psi_1 \dn \rangle \nonumber\\
  \alpha_2&=&\langle\psi_2 \up |\hat{H}_{\rm so} |\psi_2 \dn \rangle, \nonumber\\
  \beta&=&\langle\psi_1 \up |\hat{H}_{\rm so} |\psi_2 \dn \rangle, \nonumber\\
  \beta'&=&\langle\psi_1 \dn |\hat{H}_{\rm so} |\psi_2 \up \rangle.
\end{eqnarray}
To remove the off-diagonal terms mixing the $\up,\dn$ states within the same subband,
we apply the unitary transformation ${\bf B}=\underline{\underline{U}}\cdot{\bf A}$ with
\begin{equation}
  \underline{\underline{U}}=\frac1{\sqrt2}\left( \begin{array}{cccc}
    1 & -\frac{\alpha_1}{|\alpha_1|} & 0 & 0 \\
    1 & \frac{\alpha_1}{|\alpha_1|} & 0 & 0 \\
    0 & 0 & 1 & -\frac{\alpha_2}{|\alpha_2|} \\
    0 & 0 & 1 & \frac{\alpha_2}{|\alpha_2|} \
  \end{array}\right).
\end{equation}
Equation (\ref{H}) then transforms into
\begin{equation}\label{tildeH}
  \left( \begin{array}{cccc}
    \varepsilon_1-|\alpha_1| & 0 & -\gamma_1 & \gamma_2 \\
    0 & \varepsilon_1+|\alpha_1| & -\gamma_2 & \gamma_1 \\
    -\gamma_1^* & -\gamma_2^* & \varepsilon_2-|\alpha_2| & 0 \\
    \gamma_2^* & \gamma_1^* & 0 & \varepsilon_2+|\alpha_2| \
  \end{array}\right){\bf B}
 =\varepsilon {\bf B},
\end{equation}
where the off-diagonal matrix elements
\begin{equation}
  \gamma_{1,2} = \frac12 \Big[\beta\frac{\alpha_2^*}{|\alpha_2|} \pm \beta'\frac{\alpha_1}{|\alpha_1|}\Big]
\end{equation}
connect the first and second subbands. We treat these contributions to the conduction band Hamiltonian perturbatively to second order,
and obtain the following solutions of Eq. (\ref{tildeH}):
\begin{eqnarray*}
\varepsilon_1^\pm & = & \varepsilon_1 \pm |\alpha_1| \nonumber\\
&+&
\frac{|\gamma_1|^2}{\varepsilon_1\pm|\alpha_1|-\varepsilon_2 \mp |\alpha_2|}
+\frac{|\gamma_2|^2}{\varepsilon_1\pm|\alpha_1|-\varepsilon_2 \pm |\alpha_2|}, \\
\varepsilon_2^\pm & = & \varepsilon_2 \pm |\alpha_2| \nonumber\\
&+&
\frac{|\gamma_1|^2}{\varepsilon_2\pm|\alpha_2|-\varepsilon_1 \mp |\alpha_1|}
+\frac{|\gamma_2|^2}{\varepsilon_2\pm|\alpha_2|-\varepsilon_1 \pm |\alpha_1|}
\end{eqnarray*}
and
\begin{equation}\label{B1}
  {\bf B}_1^-=\left(\begin{array}{c}
    1 \\
    0 \\
    \ds \frac{-\gamma_1^*}{\varepsilon_1-|\alpha_1|-\varepsilon_2+|\alpha_2|} \\
    \ds \frac{\gamma_2^*}{\varepsilon_1-|\alpha_1|-\varepsilon_2-|\alpha_2|} \
  \end{array}\right),
\end{equation}
\begin{equation}
  {\bf B}_1^+=\left(\begin{array}{c}
    0 \\
    1 \\
    \ds \frac{-\gamma_2^*}{\varepsilon_1+|\alpha_1|-\varepsilon_2+|\alpha_2|} \\
    \ds \frac{\gamma_1^*}{\varepsilon_1+|\alpha_1|-\varepsilon_2-|\alpha_2|} \
  \end{array}\right),
\end{equation}
\begin{equation}
  {\bf B}_2^-=\left(\begin{array}{c}
    \ds \frac{-\gamma_1}{\varepsilon_2-|\alpha_2|-\varepsilon_1+|\alpha_1|} \\
    \ds \frac{-\gamma_2}{\varepsilon_2-|\alpha_2|-\varepsilon_1-|\alpha_1|} \\
    1 \\
    0 \
  \end{array}\right),
\end{equation}
\begin{equation}
  {\bf B}_2^+=\left(\begin{array}{c}
   \ds \frac{\gamma_2}{\varepsilon_2+|\alpha_2|-\varepsilon_1+|\alpha_1|} \\
   \ds \frac{\gamma_1}{\varepsilon_2+|\alpha_2|-\varepsilon_1-|\alpha_1|} \\
    0 \\
    1 \
  \end{array}\right).
\end{equation}
The eigenvectors ${\bf B}_i^\pm$ are normalized up to first order in the off-diagonal perturbation.

In the absence of {\it intra}subband (2D) terms, $\alpha_1=\alpha_2=0$, the {\it inter}subband SO interaction gives rise to spin mixing
without lifting the spin degeneracy ($\varepsilon_i^+=\varepsilon_i^-$); it only causes a spin-independent shift of the subband energies.
By contrast, if an {\it intra}subband interaction is present (or if spin degeneracy is lifted by other
means, e.g., by a magnetic field), the spin splitting is affected. For the lowest subband it is given by
$\varepsilon_1^+ - \varepsilon_1^- = \Delta \varepsilon_1$, where
\begin{eqnarray}\label{gsss}
  \Delta \varepsilon_1
&=&
  2|\alpha_1|+2|\gamma_1|^2\frac{|\alpha_2|-|\alpha_1|}{(\varepsilon_2-\varepsilon_1)^2-(|\alpha_2|-|\alpha_1|)^2} \nonumber\\
&-&
  2|\gamma_2|^2\frac{|\alpha_2|+|\alpha_1|}{(\varepsilon_2-\varepsilon_1)^2-(|\alpha_2|+|\alpha_1|)^2}.
\end{eqnarray}
To proceed further we need the explicit form of the SO Hamiltonian $\hat{H}_{\rm so}$.

\section{Rashba and Dresselhaus SO interaction in symmetric QWs}\label{sec:sym}

By folding down the $14\times 14\; {\bf k\cdot p}$ Hamiltonian for a QW grown in [001] direction in a zinc-blende crystal to a
$2\times 2$ conduction band problem \cite{PZ}, one obtains an effective SO Hamiltonian in the conduction band:
\begin{equation}\label{Hso}
\hat{H}_{\rm so} \approx \left(\begin{array}{cc} 0 & h_{\rm so} \\ h^*_{\rm so} & 0 \end{array}\right),
\end{equation}
where
\begin{equation}
h_{\rm so}=  R(z) k_- -i\lambda k_+ \frac{\de^2}{\de z^2}-i \frac{\lambda}4 (k_-^2-k_+^2)k_- \:,
\end{equation}
with
\begin{equation*} 
  \lambda = 4\frac{\sqrt2}3 PQP'
  \left(\frac1{(E_{\Delta}-\varepsilon)(E'_v-\varepsilon)}-\frac1{(E_v-\varepsilon)(E'_{\Delta}-\varepsilon)}\right)
\end{equation*}
and
\begin{eqnarray}\label{R}
  R(z)&=&\frac{\sqrt2}3 P^2\left[\frac{\de}{\de z}\left(\frac1{E_v-\varepsilon}-\frac1{E_{\Delta}-\varepsilon}\right)\right]
\nonumber\\
&+&
  \frac{\sqrt2}3 P'^2\left[\frac{\de}{\de z}\left(\frac1{E'_v-\varepsilon}-\frac1{E'_{\Delta}-\varepsilon}\right)\right].
\end{eqnarray}
Here, $k_{\pm}=\frac1{\sqrt2}(k_x\pm i k_y)$, $\varepsilon$ is the electron energy, $E_v(z)$ and $E_{\Delta}(z)$ are the
position-dependent $\Gamma_8$ and $\Gamma_7$ valence band edges, and $P = -i\frac{\hbar}{m}\langle
S|\hat{p}_x|X\rangle=\sqrt{E_p\frac{\hbar^2}{2 m}}$ is the momentum matrix element. Primed quantities correspond to the higher
lying $\Gamma_8-\Gamma_7$ conduction band and $Q$ is the momentum matrix element between the valence band and the higher
conduction band. Along with the Rashba and linear Dresselhaus terms in Eq.~(\ref{Hso}) we keep the cubic Dresselhaus term as well.
During the derivation we assumed that the variation of the band edges is small compared with the energy gaps in the material.

In symmetric structures,
due to parity conservation  the {\it intra}subband SO interaction contains only the Dresselhaus contribution,
\begin{eqnarray}\label{a1}
  \alpha_1 &= & -\frac{\lambda}{4\sqrt2} k^3 \sin(2\varphi)\; e^{-i\varphi} + \frac{D_{11}}{\sqrt2} k\; e^{i(\varphi+\frac{\pi}2)}, \\
\label{a2}
  \alpha_2 &= & -\frac{\lambda}{4\sqrt2} k^3 \sin(2\varphi)\; e^{-i\varphi} + \frac{D_{22}}{\sqrt2} k\; e^{i(\varphi+\frac{\pi}2)},
\end{eqnarray}
and the {\it inter}subband SO interaction (between the lowest two subbands) involves only the Rashba term
\begin{equation}
  \beta = \beta'^*=\frac{R_{12}}{\sqrt2} k e^{-i\varphi},
\end{equation}
where $\varphi$ is the polar angle of the in-plane vector ${\bf k}_{\prll}$ measured from the [100] direction, and $k=|{\bf k}_{\prll}|$.
Furthermore,
\begin{equation} \label{dii}
  D_{ii}=-\lambda \left\langle \psi_i(z) \left| \frac{\de^2}{\de z^2} \right| \psi_{i}(z) \right\rangle
\end{equation}
and
\begin{equation} \label{r12}
  R_{12}=\langle \psi_1(z) |R(z)|\psi_2(z) \rangle.
\end{equation}
The quantity $R_{12}$ corresponds to the coupling parameter $\eta$ derived in  Ref. \cite{egues} using an 8-band ${\bf k} \cdot {\bf p}$
model.

For small $k$ the linear term in Eqs.~(\ref{a1})-(\ref{a2}) dominates and we can approximate
\begin{equation}
\frac{\alpha_1}{|\alpha_1|}
\approx \frac{\alpha_2}{|\alpha_2|}
\approx e^{i(\varphi+\frac{\pi}2)}.
\end{equation}
Then,
\begin{eqnarray}
  \gamma_1&=&\frac1{\sqrt2} R_{12} k \cos\left(2\varphi+\frac{\pi}2\right)\\
  \gamma_2&=&-\frac{i}{\sqrt2} R_{12} k \sin\left(2\varphi+\frac{\pi}2\right),
\end{eqnarray}
and the ground state spin splitting follows from Eq.~(\ref{gsss})  as
\begin{equation}
  \Delta \varepsilon_1 \approx
  2|\alpha_1|-\frac{R_{12}^2 D_{11}}{\sqrt2 (\varepsilon_2-\varepsilon_1)^2} k^3 -
  \frac{R_{12}^2 D_{22}}{\sqrt2 (\varepsilon_2-\varepsilon_1)^2} k^3 \cos(4\varphi).
\end{equation}
The {\it inter}subband interaction results thus in an additional spin splitting proportional to $k^3$.

Next, we expand  the spin splitting that is induced by the
{\it intra}subband SO interaction. Up to order $k^3$ we obtain
\begin{equation}
  |\alpha_1|\approx \frac{D_{11}}{\sqrt2} k + \frac{\lambda}{8 \sqrt2} k^3 - \frac{\lambda}{8 \sqrt2} k^3 \cos(4\varphi),
\end{equation}
which gives the final expression for the subband splitting:
\begin{eqnarray}
  \Delta \varepsilon_1 &=&  \sqrt2 D_{11} k +
  \left(\frac{\lambda}4 -\frac{R_{12}^2 D_{11}}{(\varepsilon_2-\varepsilon_1)^2}\right) \frac{k^3}{\sqrt2} \nonumber\\
&-&
  \left(\frac{\lambda}4 +\frac{R_{12}^2 D_{22}}{(\varepsilon_2-\varepsilon_1)^2}\right) \frac{k^3}{\sqrt2} \cos(4\varphi).
\end{eqnarray}
One finds that the {\it inter}subband SO interaction produces an additional spin splitting of the same
symmetry as the {\it intra}subband cubic Dresselhaus term. We will now estimate the magnitude of
this additional contribution for GaAs parabolic QWs.

\section{Subband spin splitting in parabolic wells} \label{parabolic}

Let us consider a parabolic QW with conduction band confining potential
\begin{equation}
  V(z)=\frac12 K z^2,
\end{equation}
 resulting in the noninteracting energy spectrum
\begin{equation}
  \varepsilon_j=\sqrt{\frac{\hbar^2 K}{m^*}}\left(j-\frac12 \right), \hspace{2cm} j=1,2,\dots
\end{equation}
The first and second subband envelope functions are
\begin{eqnarray}
  \psi_1(z)& = & \sqrt[4]{\frac{2\xi}{\pi}} e^{-\xi z^2}, \\
 \psi_2(z)& = & \sqrt[4]{\frac{32\xi^3}{\pi}} z e^{-\xi z^2},
\end{eqnarray}
where $\xi=\sqrt{m^* K/4 \hbar^2}$.
Straightforward calculations give
\begin{eqnarray}
  \left\langle \psi_1 \left|\frac{\de^2}{\de z^2} \right| \psi_1 \right\rangle & = & -\xi, \label{d1}\\
 \left\langle \psi_2 \left|\frac{\de^2}{\de z^2} \right| \psi_2 \right\rangle & = & -3 \xi, \label{d2}\\
  \left\langle \psi_1 \left| z \right| \psi_2 \right\rangle & = & -\frac1{2\sqrt{\xi}} \label{d3} \:.
 \end{eqnarray}
For our parabolic well, the positional dependence of the valence band edge (the valence band potential) is
\[E_v=-\frac14 K z^2, \]
corresponding to a valence band offset VBO=0.33. For GaAs parameters ($E_g=1.42$ eV, $\Delta=0.34$ eV, $E_p=22$~eV)
Eq.~(\ref{R}) gives $R(z)\approx -\left(\frac{\de E_v}{\de z}\right) 7 \rm \AA^2$.
Using Eqs. (\ref{dii}), (\ref{r12}) and (\ref{d1})--(\ref{d3}) we then get
\begin{equation*}
  R_{12}=-\frac{(7 {\rm \AA}^2) K}{4\sqrt{\xi}} , \hspace{1cm} D_{11}=\lambda \xi,
  \hspace{1cm} D_{22}=3\lambda \xi,
\end{equation*}
and
\begin{equation*}
  \frac{R_{12}^2 D_{22}}{(\varepsilon_2-\varepsilon_1)^2}=
  \left(\frac{\Delta \varepsilon}{\frac{\hbar^2}{2 m^* {\rm \AA}^2}}\right)^2 \frac{147}{64}\lambda \sim 10^{-6} \lambda,
\end{equation*}
for $m*=0.065 m_0$ and $\Delta \varepsilon = \varepsilon_2-\varepsilon_1=40$ meV. The contribution of the {\it inter}subband SO
interaction to the spin splitting of the lowest conduction subband is six orders of magnitude weaker than that of the cubic Dresselhaus {\it
intra}subband term and thus can be completely neglected.

The spin mixing induced by the {\it inter}subband SO interaction can be estimated from Eq.~(\ref{B1}):
\begin{equation*}
  \left| \frac{\gamma_2}{\varepsilon_2-\varepsilon_1} \right|^2 \approx \frac{R_{12}^2 k^2}{2 (\Delta\varepsilon)^2}=
  \frac{49}{32}\left(\frac{\Delta \varepsilon}{\frac{\hbar^2}{2 m^* {\rm \AA}^2}}\right) k^2 {\rm \AA}^2 \sim 10^{-7},
\end{equation*}
for $k=0.01\,{\rm \AA}^{-1}$. This is seven orders of magnitude weaker than the spin mixing induced by {\it intra}subband SO interaction
and also can be completely neglected.

Similar results were obtained for GaAs symmetric rectangular QWs.

\section{Conclusions} \label{conclusions}

In this paper, we have considered the effects of SO coupling on the conduction subband states in symmetric QWs. Our work
was motivated by Ref. \cite{egues}, which discussed a SO coupling effect specific to QWs with more than one subband and
showed that it can affect the electronic and spin transport properties in some systems.

We found that although the magnitude of this {\it inter}subband SO interaction
can be comparable to that of the 2D Dresselhaus and Rashba terms, its effect on the spin splitting and spin mixing of
conduction band states is several orders of magnitude weaker since it connects states with different energies.
This is due to the fact that the spin splitting and spin mixing of conduction band states are renormalized by the
{\it inter}subband energy difference.

Therefore, if one considers system with non-degenerate subbands, one can completely
neglect the {\it inter}subband SO interaction compared to the usual 2D Dresselhaus and Rashba terms.
These findings provide an  {\it a posteriori} justification for the approach used to calculate subband splittings and spin plasmon dispersions
carried out in Ref. \cite{UllrichFlatte}. This opens the way for a comprehensive theory of collective intersubband
excitations in QWs in the presence of SCD and SO coupling.

\section*{Acknowledgments}

This work was supported by DOE Grant No. DE-FG02-05ER46213,
the Nuffield Foundation Grant NAL/01070/G, and
by the Research Fund 10024601 of the Department of Physics of the University of York.

\end{document}